\documentclass[12pt,preprint]{aastex}

\newcommand{\Rsun}{\rm $R_\odot$}
\newcommand{\msun}{\rm $M_\odot$}

\shorttitle{WIRED III: A Dusty WD+M Eclipsing Binary} 
\shortauthors{Debes et al.}
\begin{document}
\title{The WIRED Survey III: An Infrared Excess around the Eclipsing Post-Common Envelope Binary SDSS J030308.35+005443.7}
\author{John H. Debes\altaffilmark{1}, D. W. Hoard\altaffilmark{2}, Jay Farihi\altaffilmark{3}, Stefanie Wachter\altaffilmark{4}, David T. Leisawitz\altaffilmark{5}, Martin Cohen\altaffilmark{6}}

\altaffiltext{1}{Space Telescope Science Institute, Baltimore, MD 21218}
\altaffiltext{2}{Spitzer Science Center, California Institute of Technology, Pasadena, CA 91125}
\altaffiltext{3}{Department of Physics \& Astronomy, University of Leicester, Leicester LE1 7RH}
\altaffiltext{4}{IPAC, California Institute of Technology, Pasadena, CA} 
\altaffiltext{5}{Goddard Space Flight Center, Greenbelt, MD 20771}
\altaffiltext{6}{Monterey Institute for Research in Astronomy, Marina, CA 93933}

\begin{abstract}
We present the discovery with {\em WISE} of a significant infrared excess associated with the eclipsing post-common envelope binary SDSSJ 030308.35+005443.7, the first excess discovered around a non-interacting white dwarf+main sequence M dwarf binary.  The spectral energy distribution of the white dwarf+M dwarf companion shows significant excess longwards of 3\micron.  A T$_{eff}$ of 8940~K for the white dwarf is consistent with a cooling age $>$2~Gyr, implying that the excess may be due to a recently formed circumbinary dust disk of material that extends from the tidal truncation radius of the binary at 1.96~\Rsun\ out to $<$0.8~AU, with a total mass of $\sim$10$^{20}$~g.  We also construct {\em WISE} and follow-up ground-based near-infrared light curves of the system, and find variability in the $K$ band that appears to be in phase with ellipsoidal variations observed in the visible.  The presence of dust might be due to a) material being generated by the destruction of small rocky bodies that are being perturbed by an unseen planetary system or b) dust condensing from the companion's wind.  The high inclination of this system, and the presence of dust, make it an attractive target for M dwarf transit surveys and long term photometric monitoring.

\end{abstract}

\keywords{circumstellar matter--binary stars--white dwarfs}

\section{Introduction}

The WISE InfraRed Excesses around Degenerates (WIRED) survey is designed to detect infrared (IR) excesses around white dwarfs (WDs) using photometry from the WISE mission \citep[for a complete description of the WISE mission see][]{wright10}.  Dust, low mass companions, and cyclotron radiation from accreting magnetic WDs all emit at mid-IR wavelengths, providing a rich variety of sources to be discovered.  Because of the all-sky coverage of WISE, WIRED provides a more systematic and less biased search for IR excesses around WDs than those performed with targeted {\em Spitzer} observations.

In \citet[][Paper I]{wdebes10}, we characterized the IR excess discovered around GALEX 1931+0117 using WISE photometry.  In \citet[][Paper II]{wdebes11}, we cross-correlated visible, near-IR, and WISE photometry of WDs discovered in the SDSS DR7 Preliminary WD catalog \citep{kleinman11} to discover excesses around WDs due to dust disks as well as low mass stellar and substellar companions.  In the process, we have discovered a WD+dM system,SDSS J030308.35+005444.1 (hereafter SDSS J0303+0054), that shows an infrared excess above the photosphere of the M star companion.  SDSS~J0303+0054 has a $W1-W3$ color of 1.8$\pm$0.1, much redder than the average $W1-W3$=0.5$\pm$0.2 of M4--M6 dwarfs detected by WISE with $W3$ magnitudes having signal-to-noise (S/N)$>$ 5 \citep{kirkpatrick11}.  

SDSS J0303+0054 is an eclipsing post-common envelope binary first discovered as a WD+dM system in the SDSS DR4 WD catalog \citep{eisenstein06, silvestri06}.  \citet[][hereafter P09]{pyrzas09} detected significant radial velocity variations in the SDSS J0303+0054 system, as well as eclipses of the WD from optical photometric monitoring.  They determined that SDSS J0303+0054 is a DC+dM4 binary with a 3.2 hr orbit.  From the combination of radial velocities and light curve modeling they constrained the mass of the WD, assumed to have a helium dominated atmosphere, to 0.88-0.95 M$_\odot$ and the companion's mass to between 0.22-0.28 M$_{\odot}$.  The T$_{eff}$ of the WD is not well constrained due to a lack of
absorption features that might otherwise be detected if not
for the overwhelming flux of the M dwarf companion; however,
P09 estimated that it must be $<$8000~K based on spectral decomposition of the two stellar components.  

Observations at $u$, $g$, and $i$ were also conducted by \citet{parsons10}, showing almost complete eclipses of the WD at short wavelengths, a brightening of the system during eclipse of the WD at $u$ and $g$, and ellipsoidal variations in $i$.  \citet{tappert11} also found evidence for multiple H$\alpha$ components, which could be caused by a combination of activity/irradiation of the M dwarf and the accretion of material on the WD surface.  

Based on current prescriptions of orbital migration after common envelope (CE) evolution \citep[e.g.][]{schreiber03}, SDSS J0303+0054 should have left the CE with a period of $\approx$0.2 day, and slowly evolved through magnetic braking to its current orbit \citep{zorotovic10}, where the M dwarf does not overfill its Roche lobe and hence no mass transfer through the L1 point should occur in this system.  Once the companion fills its Roche lobe and mass transfer through the L1 point begins, the binary becomes a cataclysmic variable (CV) and circumbinary dust is a distinct possibility.  Some CVs have been found to show evidence for cold dust in a circumbinary disk \citep{hoard09}.  This dust could originate from condensation of gaseous or dusty material that is lost from the inner binary; for example, during classical nova outbursts, in a wind from the accretion disk and/or secondary star, or as ``spillage'' from the mass transfer process between the two stars.  Numerous single WDs show IR excesses due to dust \citep[e.g.][and references therein]{farihi10,xu11,wdebes11}, likely produced by the tidal destruction of asteroids perturbed into the WD Roche limit by remnant planetary systems \citep{debes02,jura03,bonsor11,debes11a}.  Given the large mass and relatively large cooling age of the WD in SDSS~J0303+0054 ($\sim$3~Gyr, P09, the presence of an infrared excess is an intriguing development in the study of this binary system.

We present follow-up observations of the SDSS~J0303+0054 system in \S \ref{s:obs}.  In \S \ref{s:sed} we construct a spectral energy distribution (SED) of the WD+dM in order to accurately determine the SED of the IR excess.  We construct some simple models of the excess in \S \ref{s:models}.  In \S \ref{sec:nir} we present our follow-up near-IR observations of SDSS~J0303+0054 as well as timeseries WISE photometry to determine the possible photometric variability in this binary.  and discuss the implications of our work in \S \ref{s:disc}.

\section{Observations}
\label{s:obs}
In addition to GALEX, SDSS DR7, 2MASS, UKIDSS, and WISE photometry, follow-up Near-IR observations of SDSS~J0303-0054 were obtained in clear conditions on 20 and 21 October 
2011 with the 4.2\,m William Herschel Telescope using the Long-Slit Intermediate Resolution Infrared 
Spectrograph (LIRIS; \citealt{man98}).  The follow-up data were obtained to search for photometric variability on orbital timescales and to obtain additional contemporaneous photometry of the system.  Additionally, we obtained the All Sky Single Exposure photometry\footnote[1]{http://wise2.ipac.caltech.edu/docs/release/prelim/expsup/sec1\_2.html\#singlexp\_images} and WISE 3-Band Cryo Single Exposure photometry\footnote[2]{http://wise2.ipac.caltech.edu/docs/release/allsky/expsup/sec7\_1.html} in $W1$ and $W2$ of SDSS~J0303+0054 in the {\em WISE} All-Sky and 3-Band Cryo Catalogs in order to track possible short-term variability.

On the first night of the LIRIS run, a single 9-point dither pattern of images was 
obtained in the $J$, $H$, and $K_s$ bands, with total exposure times of 18\,s in each filter.  This was
followed immediately by 10 identical and consecutive sequences at $H$, then again at $K_s$.  On 
the second night, 50 such consecutive sequences were executed at $K_s$ only, spanning approximately 
1.2 hours in total elapsed time.  On both nights, four standard star fields (ARNICA, 11 stars total;\citealt{hunt98}) were observed in 
a similar manner for photometric zero-point calibration.  The data were reduced in the standard manner, 
by subtracting a median sky from each image in the dithered stack, flat fielding (using sky flats), then 
averaging and recombining frames.  All data taken in the Ks-band filter were flux-calibrated using
the ARNICA K-band standard star photometry, and the error introduced
by this should be significantly smaller than the 5\% absolute
calibration uncertainty.

LIRIS suffers from what is known as a detector reset anomaly, which appears in certain frames as a 
discontinuous jump (in dark current) between the upper and the lower two quadrants. To remove this 
unwanted signal, after flat fielding and sky subtraction, the detector rows were collapsed into a median 
column (with real sources rejected), and subsequently subtracted from the entire two dimensional image.  
The resulting fully reduced frames exhibit smooth backgrounds, free of the anomalous gradient. 

\section{Construction of the system SED}
\label{s:sed}

\begin{figure}
\plotone{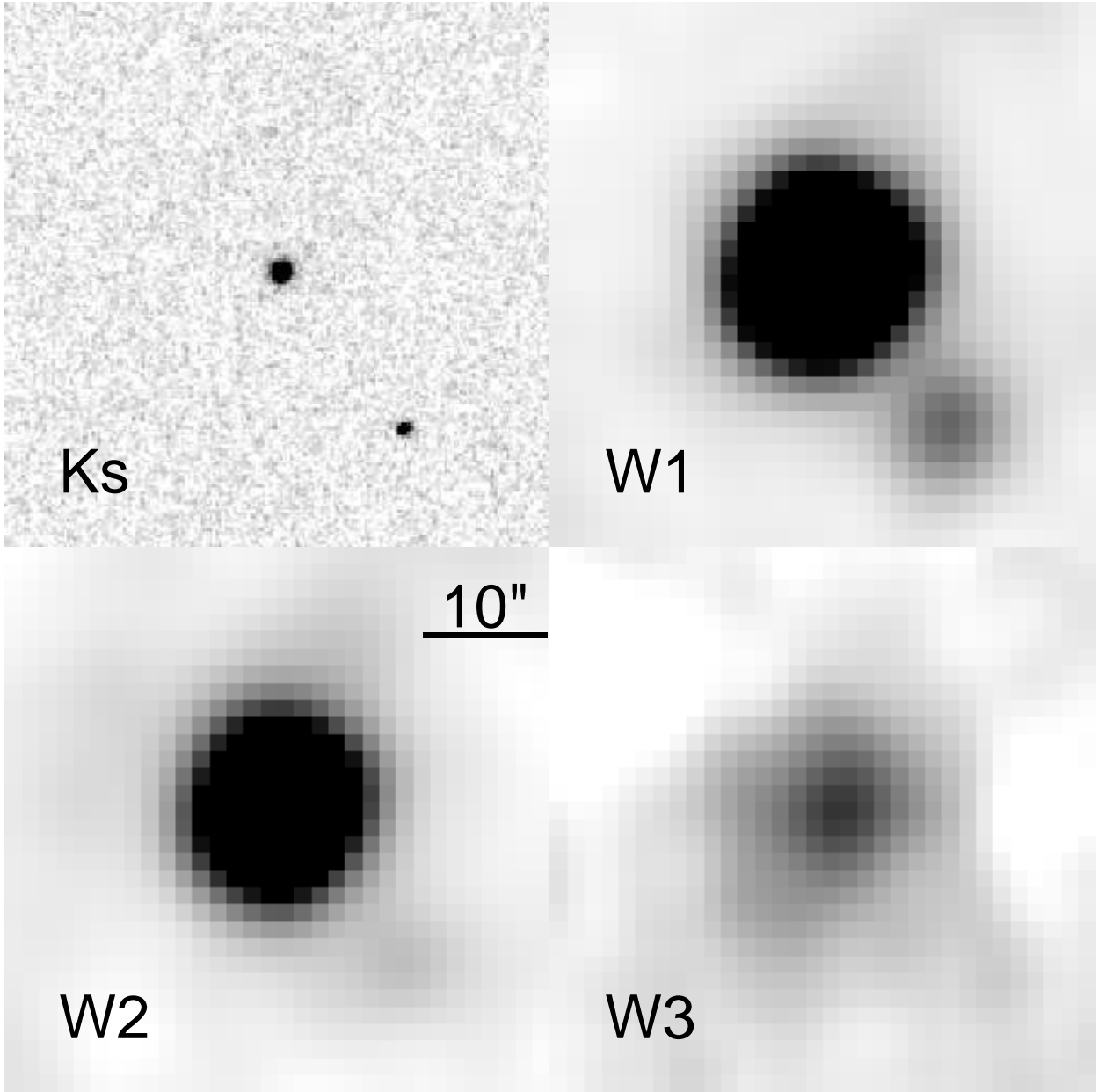}
\caption{\label{fig:f1} Infrared images of SDSS~J0303+0054 in LIRIS $K_s$-band, and the WISE $W1$, $W2$, and $W3$ bands centered on its SDSS DR 7 coordinates.  North is up and East is to the left.  Greyscale goes from 2$\sigma$ below local background level to 25$\sigma$ above the local background.}
\end{figure}

Figure \ref{fig:f1} shows LIRIS $K_s$, WISE $W1$, $W2$, and $W3$ images of SDSS J0303+0054.  It is detected with a S/N of $>$100, 43, 42, and 11, respectively.  Investigation of the PSF in each WISE band does not reveal any extension: the science target in each band appears to be consistent with the expected PSF FWHM.  Furthermore, inspection of higher spatial resolution UKIDSS, and LIRIS images show no indication of any neighboring sources, limiting contamination from comparably bright sources to angular separations $<$0\farcs35.  Table \ref{tab:phot} lists the GALEX, SDSS, near-IR, and WISE photometry for this system, the modified julian date (MJD) of each observation after a heliocentric correction and its implied orbital phase for the binary, as well as our best fit model described below.  

\subsection{Empirical M dwarf photometry}

An integral part of our SED modeling is the comparison to empirical SEDs of M dwarfs with known spectral types.  In order to construct our empirical SEDs, we gathered optical, NIR, and WISE photometry of known, bright M dwarfs.  Average optical Sloan $g-r$ colors as a function of spectral type were collected from \citet{west11}, while average $r-i$, $i-z$, and $z-J$ colors and absolute $J$ magnitudes were obtained from \citet{hawley}.  These colors were tied to 2MASS and WISE photometry collected for M, L, and T dwarfs \citep{kirkpatrick11}.  For a smooth distribution of colors to allow for interpolation between spectral types and to estimate the potential uncertainty in the SED templates, we fit low order polynomials to each color and used those fits for each color.  For the $u$ and GALEX filters, we assumed a Wien extrapolation from the $g$ filter where $F(\lambda)\propto(\lambda/0.477~\micron)^5$.

\subsection{Fitting models to the SED}
\begin{figure}
\plotone{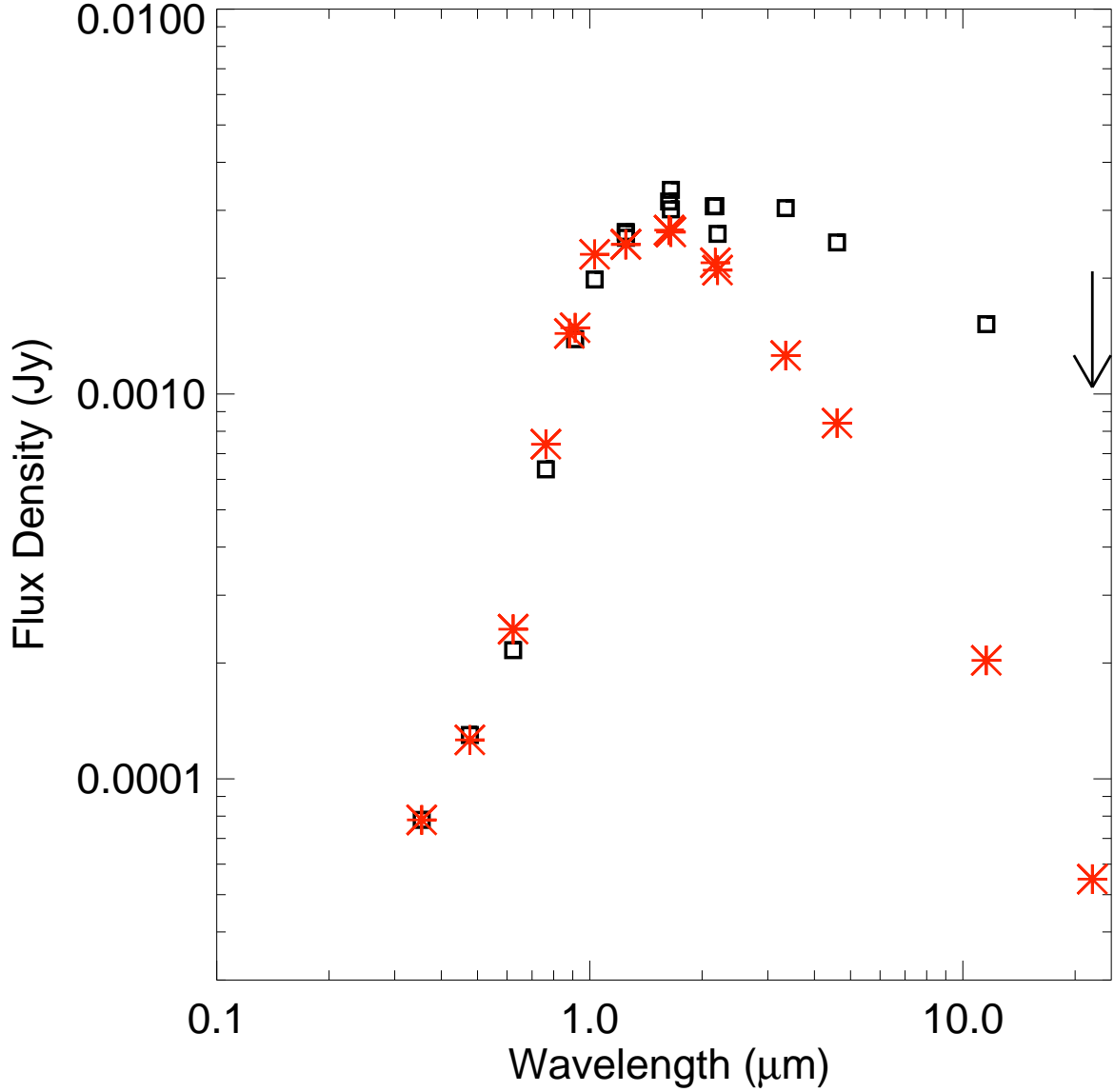}
\caption{\label{fig:f2}  SED of the SDSS J0303+0054 system from GALEX $FUV$ to $W3$, with an upper limit in $W4$.  Black squares are the observed photometry, including the near-IR photometry at $J, H, and K$ at three separate epochs.  Errors are typically smaller than the symbol size.  Overplotted with red asterisks is a best-fitting model DC+dM6 SED, with a T$_{eff}$=8940~K, $\log$ g=8.5 WD.}
\end{figure}

In order to determine the significance of the possible nature of the IR excess, one must accurately subtract off the emission from the two stellar components.  We first converted the photometry of SDSS J0303+0054 into units of flux density as in \citet{wdebes11}, and then simultaneously fit empirical SEDs of M dwarfs and model SEDs of WDs to the observed photometry.  WD SEDs were constructed from He-rich cooling models kindly provided by P. Bergeron that include the GALEX and WISE bands \citep{bergeron95,holberg06}.  We compared our models with GALEX FUV through $J$ photometry of SDSS J0303+0054, determined a median scaling between the observed and model photometry, and minimized the $\chi^2_\nu$ value to determine the best fit in $\log{g}$, T$_{eff}$, $d$ and spectral type.  The resulting best-fit parameters and their 98\% confidence intervals are shown in Table \ref{tab:fit}.  Our best fit model requires a WD with $\log{g}$=8.5, $T_{eff}$=8940~K and an M4.5 companion at a distance of 134~pc--the expected photometry is listed in Table \ref{tab:phot}.  The inferred cooling age of the WD from our models is 2~Gyr.  The resulting gravity and distance to the WD are dependent on our assumption that the companion has the same luminosity as a typical field M4.5 dwarf and is thus relatively uncertain given the inherent uncertainties in our sample of field dwarf absolute magnitudes and colors.  However, the observations of P09 quite accurately determined both WD and M dwarf radii and masses, which we use in \S \ref{s:models} and list in Tables \ref{tab:fit} and \ref{tab:adopt}.  Figure \ref{fig:f2} shows a comparison of the observed photometry and our best fitting model.  The photometry shows excess in the $H$ band, with all three sets of near-IR data showing varying levels of excess above the photosphere.  However, the three Near-IR epochs are marginally inconsistent with each other: relative to 2MASS, the LIRIS $H$ photometry is inconsistent at the 4-$\sigma$ level, and the UKIDSS $K$ photometry is inconsistent at the 3-$\sigma$ level.  These are small enough to conceivably be caused by underestimating the photometric errors, or could signal photometric variability in the system.  We address that further in \S \ref{sec:nir}.  Subtracting off the model stellar SED, we find that the $W1$, $W2$, and $W3$ bands have excesses of 22, 24, and 10 $\sigma$ above the expected combined stellar photospheres at wavelengths $>$ 3\micron.   

\section{Models for the IR excess}
\label{s:models}
We now investigate the origin of the IR excess around this system.  IR excesses around presumed single WD systems arise from a) cool, low luminosity companions, b) dust disks, or c) cyclotron emission.  Integrating over the excess and the fitted secondary SED (assuming a Rayleigh-Jeans approximation to the flux beyond 12\micron) results in $L_{IR}/L_{\star}$=0.3.  The lack of any evidence for strong magnetic fields from the WD suggests that it is unlikely to be due to cyclotron emission (e.g. as in the magnetic CV EF Eri, see \citet{hoard07b}).  The fact that the excess is brighter than the M companion at long wavelengths rules out a tertiary low-luminosity companion as the cause of the excess.  We instead examine if a circumbinary, optically thin, dust disk can qualitatively match the observed excess.  The large emitting surface area implied by the relatively large $L_{IR}/L_{\star}$ requires a circumbinary, rather than a circum-primary or circum-secondary disk--the implied radius of such a disk is larger than the orbital separation of the binary and thus would not be stable around either component \citep{lubow94}.  
In order to model the SED of the two stellar components and a circumbinary dust disk, we use the procedure developed to model IR emission from circumbinary dust observed around the CV V592 Cas \citep{hoard09}.  The model is designed to predict emission from several components in a CV, including an optically thick accretion disk and an optically thin circumbinary disk \citep{hoard07b,brinkworth07}.  In the case of SDSS J0303+0054 there is no accretion disk and so the only constituents are the WD, M companion, and the dust.

The adopted fundamental parameters of the WD and M companion that we use in our model are given in Table \ref{tab:adopt}.  These values are consistent, to within our fitting uncertainties, to the mass and gravity we derived independently for the WD from the SED alone and are consistent with the constraints placed by P09 from spectral decomposition and light curve analysis.  This combination of parameters gives a statistically indistinguishable $\chi^2_\nu$ from our best fitting model. 

\begin{figure}
\plotone{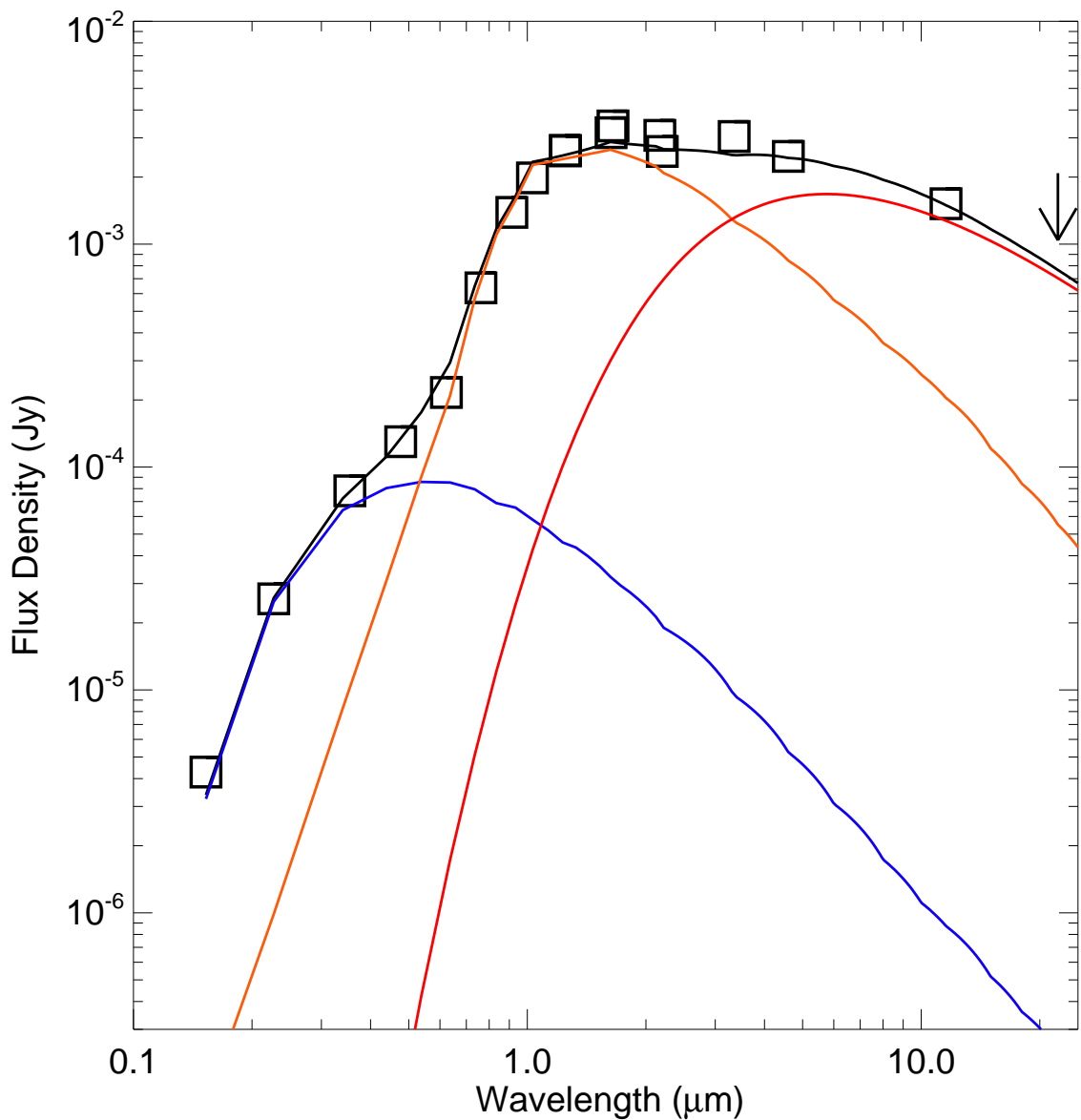}
\caption{\label{fig:f3}  Comparison of SDSS J0303+0054's photometry to our model circumbinary dust disk.  The black line is the combination of the WD (blue line), M4.5 companion (orange line), and circumbinary dust disk (red line) consisting of 1\micron\ grains arranged in a geometrically thin, optically thin disk with an inner radius of 1.96~\Rsun\ and inclination co-planar to the best fit inclination from P09 (80$^\circ$-84$^\circ$).  The outer radius of the disk is not well constrained by our models, but may extend as far as 0.8~AU.  The M dwarf companion has a luminosity an order of magnitude greater than the WD.}
\end{figure}

Using these assumptions, our model for the circumbinary disk is shown in Figure \ref{fig:f3} compared only to our roughly contemporaneous LIRIS  and WISE data.  Because of the small number of photometric points, we do not constrain a formal best fit--given the significant degeneracies present in SED modeling, we seek only to obtain a qualitative match to the data to determine whether the circumbinary disk hypothesis is plausible.  In general, our simple model reproduces the observed excess, starting at $H$ and extending out to 12~\micron.  The model underpredicts the $W1$ point, which could be due to structure in the disk that deviates from the basic structure we assumed.  The inner edge of a circumbinary disk is defined by the innermost stable orbit before tidal forces from the binary dominate a dust grain's orbit and ejects or accretes the dust grain \citep{lubow94}.  Therefore, we assume a tidal truncation radius of 218 R$_{\rm WD}$ (=1.96~\Rsun, 1.7 times the binary separation as described in \citet{hoard09}), with an inner radius temperature of 1400~K , and an outer temperature of 50~K (at 18500 R$_{\rm WD}$=0.77~AU)and a disk co-planar with the binary system.  The inferred mass of the dust is 3$\times10^{20}$~g assuming 1$\micron$ silicate grains.  If micron-sized silicate grains were present, however, one might also  expect the $W3$ band photometry to be brighter due to a strong silicate feature at $\sim$10~\micron.  This could be accounted for by assuming a smaller outer radius to the disk, but at this time a specific composition for the dust is not constrained by the data, but is used to infer a rough order of magnitude to the mass of the dust.  Furthermore, 10~\micron\ silicate features are not observed in dust around CVs \citep{brinkworth07}, implying that the dust is larger than 1\micron.  Assuming the dust all originated from a single body composed solely of silicates, the radius would be 21~km.

\section{Near-IR LIRIS and WISE Lightcurves}
\label{sec:nir}

Aperture photometry of all science targets and standard stars in our LIRIS observations was performed using an $r=3''.75$ aperture 
radii and sky annuli of $5''-7''.5$ in size, including extinction corrections.  The derived $JHK$ zeropoints
for both nights agree to within 0.01\,mag, but the scatter in the 11 standard stars each night was 0.05\,mag.
All photometry was calibrated using the $J, H$, and $K$-band system of the standard star data.  Absolute fluxes for 
SDSS~J0303+0054 were derived using the derived zeropoints for each night, and also by using 2MASS photometry
for 5 relatively bright comparison stars in the field.  The photometry thus derived for SDSS~J0303+0054 using both
methods agrees very well, within 0.02\,mag.                          

The comparison stars were also used to search for
variability in time-series relative photometry in the $K$ band.  The flux ratio of SDSS~J0303+0054 relative to the comparison stars was
calculated for all photometry resulting from each imaging sequence,
with a variance weighted-average of each set of resulting magnitudes for SDSS~J0303+0054 to create a final light curve of the system.  The heliocentric julian date was calculated from the midpoint of each photometric observation and converted to a phase using the ephemeris published in P09.  Based on the uncertainty in the period, the phases we calculate should be accurate to within $<$10$^{-2}$ or about 83 seconds.
  
WISE observed SDSS~J0303+0054 over the course of 1 day on 29 January 2010 and another day on 06 August 2010 for a total of 23 visits.  Each visit had a detection of the system, with S/N of 30 and 10 in the $W1$ and $W2$ bands, respectively.  We phased the data in the same manner as our $K$-band data.

\begin{figure}
\plotone{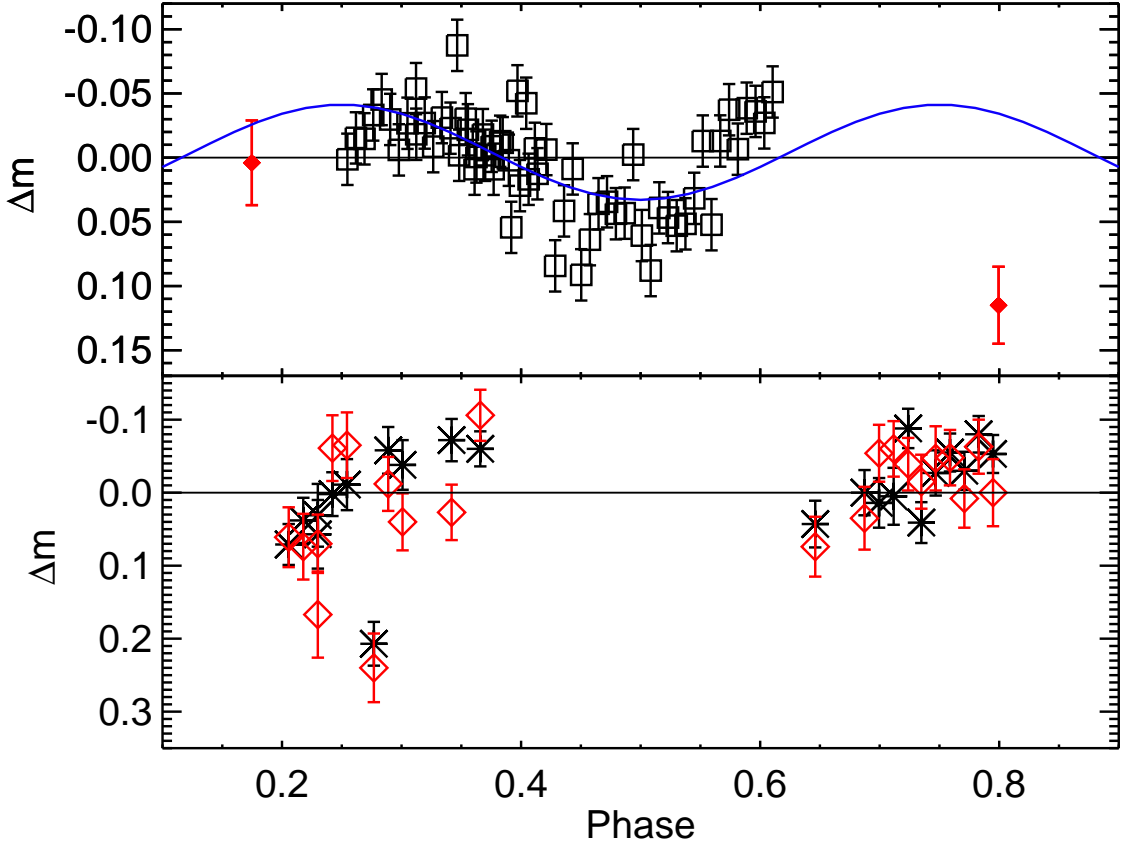}
\caption{\label{fig:f4}  (Top) $K$ light curve ($\Delta$ m=m-13.34) of SDSS~J0303+0054 phased to the ephemeris of P09.  Approximate uncertainties in the phasing should systematically be $\pm$0.01.  The diamond overplotted at phase=0.8 is the epoch 2005 UKIDSS $K$ photometry, and the diamond at phase=0.17, the epoch 2000 2MASS $K$ photometry (similar phasing uncertainties, for more discussion see \S \ref{sec:nir}).  Variability in the $K$-band brightness is evident, and we overplot a sinusoid to the data assuming a period half that of the orbital period, an amplitude of 37~mmag, and zeropoint at 0, consistent with the ellipsoidal variations observed at shorter wavelengths.  This is meant to guide the eye, rather than represent a true fit to the data.  (Bottom) WISE $W1$ (black asterisks) and $W2$ (red diamonds) photometry from 2010 and 2011 ($\Delta m=m-m_{o,W1,W2}$) phased to the same ephemeris.  There may be tentative variability in the data.}
\end{figure}

Figure \ref{fig:f4} shows the resulting $K$-, $W1$-, and $W2$-band light curves , for SDSS~J0303+0054.  In both plots, $\Delta$m is negative when the system is brighter and positive when it is dimmer.  We find that the lightcurve in the $W1$ and $W2$ photometry shows tentative evidence of variability, especially in a minor linear trend at phase$\sim$0.2, but the scatter in the data is large compared to this possible trend. The $K$ band lightcurve is clearly variable: the system varies at the level of 8\% peak-to-trough.  Based on our models, we expect the M dwarf to contribute close to 82\% of the total flux at $H$ and $K$ and 50\% (41\%) of the flux at $W1$ ($W2$).  

\citet{parsons10} found similar variations in $i$-band lightcurves with an amplitude of  10\% peak-to-trough and interpreted them to be due to ellipsoidal variations due to tidal deformation of the M dwarf by the WD.  In $i$-band, the M dwarf should contribute 80\% of the flux in the system, comparable to $H$ and $K$.  If both $H$ and $K$ vary by $\approx$10\% during an orbital period, this could account for some of the variation observed in the near-IR photometry.  The phase of our $H$-band observations overlaps with the minimum we observe at $K$, potentially explaining the dimmer value compared to 2MASS and UKIDSS.  We have also overplotted the calculated phases for the 2MASS and UKIDSS $K$ photometric data.  The 2MASS $K_s$ photometry falls roughly at a phase in the orbit where the M dwarf is expected to be close to an inflection point of the lightcurve and close to its median brightness.  If we extrapolate the behavior of the LIRIS lightcurve to the orbital phase of the UKIDSS photometry, we would expect the UKIDSS $K$ magnitude to be $\approx$13.31, almost 0.2 mag brighter than what was observed in 2005.  If the UKIDSS photometry is truly dimmer, this would imply variability in the amount of dust in the system (or of the M-dwarf) on a timescale much longer than the orbital period.  Observations of a full orbital phase in the $K$-band, as well as long term monitoring of this system, are necessary to confirm these unresolved issues.

The variability at $K$ is in phase with the ellipsoidal variations observed at $i$ and consistent to what one would expect based on the tidal distortion of the M star from the white dwarf.  Based on Roche geometry calculations neglecting limb darkening and using the mass and radius values we adopted in our circumbinary dust model, we would expect ellipsoidal variations peak-to-trough of 12\% (which scales to 10\% if the secondary contributes 84\% of the flux at $K$).  The shape of the lightcurve in Figure \ref{fig:f4} deviates slightly from a pure sine function which is also seen at shorter wavelengths \citep[P09][]{parsons10}.   We further discuss our results in the context of all the lightcurves for this system in \S \ref{s:disc}.

\section{Discussion}
\label{s:disc}

The WISE observations of SDSS J0303+0054 represent the first likely detection of dust in orbit around a non-interacting post-common envelope binary.  If the dust disk is aligned with the orbit of the binary, SDSS~J0303+0054 could be a fruitful target for high cadence near-IR transit searches.  Our photometry was sensitive to transits of depths $>$ 50 mmag and durations $>$3 minutes.  With the same photometric accuracy and sampling at $<$ 30~sec, planets with radii of $\sim$5~R$_\oplus$ could be detected if they orbited at the inner edge of the disk and transited with durations of $\sim$1 minute.  

The presence of ellipsoidal variations deviating from pure sinusoidal behavior, now seen at $K$ with our new observations, also opens up the possibility that the structure of the circumbinary disk can be directly probed by detailed analysis of long wavelength time series photometry.  The circumbinary disk is not symmetrically heated throughout an orbital period, and if there exist significant azimuthal density or compositional variations, these could be detected as asymmetries in the lightcurve.  Since the disk contributes 15\% of the light at $K$, such variations could be detected at the few percent level.

One possibility is that SDSS~J0303+0054 is the binary equivalent of isolated WD dusty disks, which most likely form from the tidal disruption of rocky bodies a few tens of kilometers in radius.  Given the relatively long cooling age of this system compared to other dusty WDs and the inferred T$_{eff}$ of the WD, this would be comparable to the coldest recorded dusty single WD observed, G166-58 \citep{farihi10}.  The amount and location of dust present at SDSS~J0303+0054 would not be detectable around a single cool WD.  The presence of the M companion's additional luminosity, which far exceeds that of the WD, enhances the detectability of the circumbinary dust.  

If the dust is caused by a tidally disrupted asteroid, then SDSS~J0303+0054 is reminiscent of Kepler-16b, which contains a 0.6\msun\ primary, a 0.2 \msun\ companion, and a Saturn-mass companion in a circumbinary orbit \citep{doyle11}.  Such a binary, with a 41 day orbit, would eventually evolve into a common envelope binary.  The remains of the circumbinary planetary system could feed a similar dust disk to that seen around SDSS~J0303+0054.  However, several potential problems exist for such a system.  Firstly, the Poynting-Robertson drag timescale might be prohibitively short for such a disk.  The P-R drag timescale can be approximated by \citep{hansen06}:

\begin{equation}
T_{\rm PR}=100 \left(\frac{s}{1\micron}\right)\left(\frac{\rho_s}{3 \rm{g~cm}^{-3}}\right)\left(\frac{r}{10^{11} \rm{cm}}\right)^2\left(\frac{L_\star}{10^{-3}L_\odot}\right)^{-1} \rm{yr},
\end{equation}

where $L_\star$ is the luminosity of the star, $s$ and $\rho_s$ are the average grain size and density respectively, and $r$ is the distance from the central star.  As noted previously, the M dwarf dominates the luminosity of the system.  Assuming a bolometric luminosity of 5$\times$10$^{-3} L_\odot$ \citep{rebassa07},  and assuming a minimum distance of the disk from the M dwarf of 0.76 \Rsun\, the PR timescale can have a range of values since the dust orbits the binary's center of mass.  At the closest distance to the M dwarf, the P-R timescale is 6~yr, while at the furthest distance away from the dust disk inner edge the P-R timescale can be as long as 100~yr.  The relatively short P-R timescale implies that dust must be replenished at a rate of 5$\times$10$^{19}$ g/yr or 2$\times$10$^{12}$ g~s$^{-1}$.  The minimum replenishment rate, corresponding to the longer P-R timescale, would be  sixteen times smaller.  This rate of dust production is high compared to the inferred accretion rates of dust onto single WDs, but could be reconciled if the radius of a typical dust particle is larger than 1\micron.  

The mechanism for creating the dust in the asteroidal scenario might be significantly different than for single WDs.  It is unclear how an asteroid could tidally disrupt at the circumbinary disk inner radius, which is much larger than the Roche disruption radii of either the WD or the M dwarf.  If the dust is from tidally disrupted bodies, multiple asteroids could have similar periastra, and the dust forms from multiple collisions of these objects.  The accretion rate would be lower if the disk were optically thick, since P-R drag could not act on all of the dust present.  Assuming an inclination of 82$^\circ$, an inclination consistent with the results of P09, the lack of reddening toward the system constrains the vertical height of the disk to be $<$ 0.1~R$_\odot$.

Alternatively, low levels of mass transfer that might be present in this system could account for the presence of a dust disk as well.  The combined photometric and spectroscopic behavior of the SDSS~J0303+0054 system suggests low levels of wind material being accreted by the WD.  The $u$ and $g$ light curves of the system show a brightening of the system coincident with the WD's eclipse.  \citet{parsons10} interpreted this excess flux as due to the side of the M-dwarf facing away from the WD being brighter.  However, this is contradicted by the $\sim$96\% depth of eclipse seen in the $u$ band, despite a 10-20\% increase in flux leading up to the eclipse.  If the M companion were responsible for the excess, the eclipse would be shallower than that predicted by the ratio of the M dwarf optical flux to the total flux from the system.  Based on our best fit model, the eclipse depth should be $\sim$90\% in $u$.  This difference suggests our model over-predicts the flux from the M dwarf, or under-predicts the flux from the WD.  We instead interpret the brightening, which covers about half the orbit, as due to the side of the WD facing the secondary being hotter due to broad wind accretion on that hemisphere.  The presence of multi-component H$\alpha$ emission lines \citep{tappert11} are also indicative of accretion from the secondary wind.  

Similar to CVs with infrared excesses, this wind accretion may provide enough ``spillage'' to create a low mass dust disk.  Using the above P-R drag mass loss rate as an estimate of dust destruction, we can infer the rough mass injection required to obtain a steady state dusty disk.  If we assume that all of the metals in the secondary's wind condense into dust, and that half of the wind must contribute to the ``spillage'', this suggests that the M dwarf must be losing mass at a rate of 5$\times10^{21}$ g yr$^{-1}$ with a minimum rate an order of magnitude lower for the longer P-R drag and assuming roughly solar metallicity.  This could be uncertain by a factor of 10 if the disk is radially optically thick--this can slow the destruction of dust grains if some of the dust is shadowed.  If we compare this to the mass loss rate of other M dwarf winds  that accrete onto WDs \citep{debes06}, we find that the M dwarf in SDSS~J0303+0054 must be losing mass at a maximum rate $\sim$100-1000 times that of typical M dwarfs and more than the solar wind mass loss rate\citep{debes06,tappert11b,pyrzas12}.  If the M dwarf is tidally distorted, as required by the presence of ellipsoidal variations, then this could enhance the mass loss in its wind.  The escape velocity at the secondary's surface is smaller than the escape speed of the binary system, which could naturally trap a majority of the wind particles in stable orbits around the binary.

Neither scenario completely explains the origins of this unusual system.  If there is significant dust production from low levels of mass transfer or from remnant planetary systems, it is possible that a significant number of other PCEBs contain dust as well.  With over 1000 candidate WD+M systems from the WIRED survey, a more careful analysis of that population may yield other dusty PCEBs, and hopefully lead to a resolution to the origin of the dust, and how it fits into the evolution of compact binaries. 


\acknowledgements
This work is based on data obtained from: (a) the Wide-field Infrared
Survey Explorer, which is a joint project of the University of
California, Los Angeles, and the Jet Propulsion Laboratory (JPL),
California Institute of Technology (Caltech), funded by the National
Aeronautics and Space Administration (NASA); (b) the Two Micron All
Sky Survey (2MASS), a joint project of the University of Massachusetts
and the Infrared Processing and Analysis Center (IPAC)/Caltech, funded
by NASA and the National Science Foundation; (c) the SIMBAD
database, operated at CDS, Strasbourg, France; and (d) the NASA/IPAC
Infrared Science Archive, which is operated by JPL, Caltech, under a
contract with NASA.  M.C. thanks NASA for supporting his participation
in this work through UCLA Sub-Award 1000-S-MA756 with a UCLA FAU 26311 to 
MIRA.

\bibliography{wired}
\bibliographystyle{apj}

\begin{deluxetable}{ccccc}
\tablecolumns{5}
\tablewidth{0pc}
\tablecaption{\label{tab:phot} SDSS J0303+0054 Photometry}
\tablehead{
\colhead{Filter} & \colhead{m} & \colhead{Model m} & \colhead {Heliocentric MJD\tablenotemark{a}} & \colhead{Orbital Phase\tablenotemark{b}}}
\startdata
$FUV\tablenotemark{c}$ & 22.32$\pm$0.14 & 22.57 & 54989.214 & 0.24 \\
$NUV\tablenotemark{c}$ & 20.38$\pm$0.03 & 20.36 & 54989.214 & 0.24\\
$u\tablenotemark{c}$ & 19.17$\pm$0.03 & 19.17 & 52243.005  & 0.87 \\
$g\tablenotemark{c}$ & 18.61$\pm$0.01 & 18.62 & 52243.005 & 0.87 \\
$r\tablenotemark{c}$ & 18.06$\pm$0.02  & 17.90 & 52243.005  & 0.87 \\
$i\tablenotemark{c}$ & 16.88$\pm$0.01  & 16.70 & 52243.005 & 0.87 \\
$z\tablenotemark{c}$ & 16.04$\pm$0.02 & 15.94 & 52243.005 & 0.87 \\
$Y_{\rm UKIDSS}$ & 15.02$\pm$0.03  & 14.83 & 53650.576 & 0.92 \\
$J_{\rm 2MASS}$  & 14.46$\pm$0.03  & 14.50 & 51789.319 & 0.17 \\
$J_{\rm UKIDSS}$ & 14.41$\pm$0.03  & 14.46 & 53650.596 & 0.07\\
$J_{\rm LIRIS}$ & 14.42$\pm$0.05  & 14.50 & 55846.134 & 0.34 \\
$H_{\rm 2MASS}$  & 13.70$\pm$0.03 & 13.94 & 51789.319 & 0.17 \\
$H_{\rm UKIDSS}$ & 13.77$\pm$0.03 & 13.92 & 53650.541 & 0.67 \\
$H_{\rm LIRIS}$ & 13.85$\pm$0.02 & 13.94 & 55846.149 & 0.46\\
$K_{s,2MASS}$ & 13.34$\pm$0.03 & 13.68 & 51789.319 & 0.17 \\
$K_{UKIDSS}$ & 13.46$\pm$0.03 & 13.66 & 53650.559  & 0.80 \\
$K_{s,LIRIS}$ & 13.31$\pm$0.02 & 13.68 & 55846.139 & 0.39 \\
$W1$ & 12.51$\pm$0.02 & 13.44 & 55225.936 & 0.22\\
...       &  12.451$\pm$0.02 & ...      & 55415.320 & 0.78 \\
$W2$ & 12.10$\pm$0.03 & 13.24 & 55225.936 & 0.22 \\
... & 12.065$\pm$0.03 & ... & 55415.320 & 0.78 \\
$W3$ & 10.70$\pm$0.10  & 12.86 & 55225.936 & 0.22 \\
... & 10.79$\pm$0.11 & ... & 55415.320 & 0.78 \\
$W4$ & $>$9.00 & 12.92 & 55225.936 & 0.22 \\
\enddata
\tablenotetext{a}{Where multiple observations exist, a mean value is calculated}
\tablenotetext{b}{Phase calculated from ephemeris of P09}
\tablenotetext{c}{AB magnitude, rather than Vega magnitude}
\end{deluxetable}

\begin{deluxetable}{ccc}
\tablecolumns{3}
\tablewidth{0pc}
\tablecaption{\label{tab:fit} SDSS J0303+0054 Fitted Fundamental Parameters}
\tablehead{
\colhead{Fundamental Parameter} & \colhead{Value} & \colhead{P09 Constraints}}
\startdata
WD $\log{g}$ & 8.5$^{+0.5}_{-1.0}$ & 8.47-8.6 \\
WD Mass (M$_\odot$) & 0.9$^{+0.3}_{-0.6}
$ & 0.88-0.95 \\
WD T$_{eff}$ (K) &  8940$^{+820}_{-400}$ & $<$8000 \\
WD Cooling Age (Gyr) &  2.0$^{+0.8}_{-1.6}$  & ... \\
Companion Spectral Type & M4.5$^{+1.0}_{-1.5}$ & M4-M5.5 \\
$\chi^2_\nu$ & 0.7 & ... \\
Inferred Distance (pc) & 134$^{+206}_{-63}$ & ... \\
\enddata
\end{deluxetable}

\begin{deluxetable}{ccc}
\tablecolumns{2}
\tablewidth{0pc}
\tablecaption{\label{tab:adopt} SDSS J0303+0054 Adopted Fundamental Parameters}
\tablehead{
\colhead{Fundamental Parameter} & \colhead{White Dwarf} & \colhead{M dwarf}}
\startdata
Mass (M$_\odot$) & 0.89 & 0.25 \\
Radius (R$_{\odot}$) & 0.009 & 0.26\\
Inferred T$_{eff}$ & 8960 & 3020 \\
Inferred Distance (pc) & 135 & ... \\
Luminosity (L$_\odot$) & 4.6$\times$10$^{-4}$ &  5.0$\times$10$^{-3}$\\
\enddata
\end{deluxetable}

\end{document}